\documentclass[12pt]{article}
\usepackage{amsmath, amsthm, amsfonts, amssymb, mathrsfs, graphicx}

\begin{document}

\title{Reformulating the NMR Quantum Mechanics Fundamental Aspects:\\
Spin$\tfrac 12$ Evolution in Magnetic Field Gradients}
                                              
\author{ $^{a}$Tarek Khalil
\footnote{E-mail address: tkhalil@ul.edu.lb}\\
and\\
$^{b}$Daniel Grucker
\footnote{ E-mail address: d.grucker@unistra.fr}}

\maketitle
\begin{center}
$^{a}$ Department of Physics, Faculty of Sciences(V), Lebanese  University, Nabatieh, Lebanon\\
$^{b}$ Laboratoire Icube, UMR 7357 CNRS/UdS, Universit\'e de Strasbourg, France\\
\end{center} 

\begin{abstract}
\noindent
Magnetization of a spin$\tfrac 12$ set is determined by means of their individual wave function. The theoretical treatment based on the fundamental axioms of quantum mechanics and solving explicitly Schr\"{o}dinger equation gives the evolution of spin$\tfrac 12$ system driven by magnetic fields. In this work we consider the energy of the spin system interacting with magnetic fields and all the other parts of the energy as a constant reservoir. Solving this complete Hamiltonian can explain the measured sign change of nuclear spin noise spectra compared to conventional NMR experiments and the possibility to make NMR images without radio frequency magnetic fields. The possibility to make NMR images without RF magnetic fields is an indication that entangled spin$\tfrac 12$ states can be manipulated by magnetic field gradients, opening a new way to perform quantum computation by NMR.

\end{abstract}


\newpage 
\section*{Introduction}
The work of Norbert M\"{u}ller and Alexej Jerschow  \cite{ Muller2006} on nuclear spin noise imaging has reactivated the fundamental idea given by Felix Bloch \cite{Bloch1946} in 1946 that N nuclei with magnetic moment $\mu$ can be measured, even in absence of magnetic field, because of statistically incomplete cancellation. A review on nuclear spin noise which corresponds to this idea can be found in reference \cite{ Muller2013}. Generally, most of quantum systems are treated  in contact with an environment which may be of different types. These systems correspond to a so called open quantum systems \cite{riv1}. The aim of this work is to give the frame for a quantum mechanical description of the evolution of a system consisting of an ensemble of $N$~identical time dependent spin$\tfrac 12$ moving particles which experience the presence of an external perturbation, as for instance main static and gradient magnetic fields with or without radio frequency fields. In other words we deal with spin$\frac 12$  system driven by external time-dependent perturbations. Solving step by step the Schr\"{o}dinger equation and defining the complex susceptibility of one spin$\tfrac 12$ as the product of the amplitude of one spin state time the complex conjugate of the other state permits to retrieve the expression of the magnetization, the absence of a significant induced current by an ensemble of spin$\tfrac 12$ at equilibrium, the existence of an induced current after a RF magnetic field pulse in magnetic resonance experiments and the existence of the spin noise. This derivation should be helpful for new magnetic resonance imaging sequences or for developing quantum computing with  spin $\tfrac 12$ using magnetic field gradients, in order to create qubits not only by the different local magnetic fields existing in a molecule. The theoretical description will be tested with the main results of NMR experiments. 
\section{Theoretical description}
\subsection{One spin$\tfrac 12$ in a static magnetic field}
For one spin$\tfrac 12$ in a $\boldsymbol{\vec{B}_0}$~magnetic field, it is well known that the potential part of the Hamiltonian is:
\begin{align}\label{eq1}
\begin{split}
\mathscr{H}_1&=-\boldsymbol{\vec{\mu}} \cdot \boldsymbol{\vec{B}_0}
=-\gamma  \boldsymbol{\vec{I}}_1\cdot \boldsymbol{\vec{B}_0},
\end{split}
\end{align}
where $\gamma$ is the gyromagnetic ratio of the spin and $\boldsymbol{\vec{I}}_1$ is the spin that verifies the following equation:
\begin{equation}
\boldsymbol{\vec{I}}_1=\frac 12\hbar\boldsymbol{\vec{\sigma}}, 
\end{equation}
with $\boldsymbol{\vec{\sigma}}$ a vector defined by the following Pauli matrices:
\begin{align}
\sigma^X=\begin{pmatrix}
            0  & 1  \\
	    1  & 0  \\
           \end{pmatrix},&&
\sigma^Y=\begin{pmatrix}
            0  & -i  \\
	    i  & \phantom{-}0  \\
           \end{pmatrix},&&
\sigma^Z=\begin{pmatrix}
            1  & \phantom{-}0  \\
	    0  & -1  \\
           \end{pmatrix}.
\end{align}
The matrix notation of equation~\eqref{eq1} is given by:
\begin{align}\label{eq2}
\mathscr{H}_1 &= -\tfrac 12 \gamma \hbar 
\begin{pmatrix}
B_Z  & B_X-iB_Y  \\
B_X+iB_Y  & -B_Z  \\
\end{pmatrix}.
\end{align} 
By convention, the vector~$\boldsymbol{\vec{B}_0}=\left(B_X,B_Y,B_Z\right)$ is placed in the $Oz$-direction, that is $\boldsymbol{\vec{B}_0}=\left(0,0,B_0\right)$ and the $xOy$-plane is called the transverse plane. Therefore the matrix form of the Hamiltonian reduces to:
\begin{align} \label{H1}
\mathscr{H}_1 &= -\tfrac 12 \gamma \hbar 
\begin{pmatrix}
B_0  & \phantom{-}0  \\
0  & -B_0  \\
\end{pmatrix}.
\end{align}
The two eigenvalues of the Hamiltonian~$\mathscr{H}_1$, which give the energy of the quantum states, are: $E_- = -\frac 12 \gamma \hbar B_0$ and $E_+ =  \frac 12 \gamma \hbar B_0$ which have the following two corresponding eigenvectors:
\begin{align}
\left\lvert-\right\rangle= 
\begin{pmatrix}
0 \\
1 \\
\end{pmatrix},&&
\left\lvert+\right\rangle= 
\begin{pmatrix}
1 \\
0 \\
\end{pmatrix}.
\end{align}
We can describe the wave function as:
\begin{align}
\label{wavefunction}
\bigl\lvert\psi_1(0)\bigr\rangle = r_1 e^{i\phi_1} \left\lvert-\right\rangle + r_2 e^{i\phi_2}\left\lvert+\right\rangle
=
\begin{pmatrix}
r_2e^{i \phi_2} \\
r_1e^{i \phi_1} \\
\end{pmatrix}
\text{with} && r_1^2+r_2^2=1
\end{align}
The 1 subscript corresponds to the low energy or ground level.
The time evolution of this quantum system (far from the speed of light) is given by the Schr\"{o}dinger equation: 
\begin{align}
\frac{i\hbar \partial\bigl\lvert\psi_1(t)\bigr\rangle}{\partial t}&= \mathscr{H}\bigl\lvert\psi_1(t)\bigr\rangle. 
\end{align}
If we write
\begin{align}
\bigl\lvert\psi_1(t)\bigr\rangle&=
\begin{pmatrix}
  x_2(t)\\
  x_1(t)\\
\end{pmatrix},
\end{align}
then we have the following differential equations:
\begin{align}\label{Dif1}
\begin{split}
 i\hbar
 \begin{pmatrix}
  \dot{x}_2(t)\\
  \dot{x}_1(t)
 \end{pmatrix}&=-\tfrac 12 \gamma \hbar 
\begin{pmatrix}
B_0  &\phantom{-} 0  \\
0  & -B_0  \\
\end{pmatrix}\cdot
\begin{pmatrix}
x_2(t)\\
x_1(t)\\
\end{pmatrix}\\
&=\tfrac 12\hbar\begin{pmatrix}
   \phantom{-} \omega_0 x_2(t)\\
   -\omega_0 x_1(t)\\
  \end{pmatrix},
\end{split}\\
\intertext{where} 
\omega_0&=-\gamma B_0 
\end{align}
 where $\omega_0$ is the resonance or Larmor frequency. The following time dependent wave function is the obvious solution of this system of equations:
\begin{equation}
\label{1spinB0}
\bigl\lvert\psi_{1}(t)\bigr\rangle= 
\begin{pmatrix}
r_2 e^{-i\bigl(\tfrac{\omega_0 t}{2}-\phi_2\bigr)}\\
r_1 e^{+i\bigl(\tfrac{\omega_0 t}{2}+\phi_1\bigr)}\\ 
\end{pmatrix}
=
\begin{pmatrix}
 e^{-i\tfrac{\omega_0 t}{2}} & 0\\
 0 & e^{+i\tfrac{\omega_0 t}{2}}\\ 
\end{pmatrix}
\bigl\lvert\psi_{1}(0)\bigr\rangle
=A(\omega_0, t) \bigl\lvert\psi_{1}(0)\bigr\rangle
\end{equation}
which defines the spin evolution time operator $A(\omega_0, t)$ of spin$\tfrac 12$                     in a static magnetic field. We notice that $A$ is a $4\pi$ period function and that the $\bigl\lvert\psi_{1}(0)\bigr\rangle$ is the wave function given by equation \eqref{wavefunction}.
\subsection{The effect of a RF magnetic field}
In magnetic resonance experiments, a transition between the two states $\left\lvert-\right\rangle$ and $\left\lvert+\right\rangle$ is obtained by a $B_1$~magnetic field rotating in the transverse plane, the $(xOy)$-plane.
\noindent This field is generated by an electromagnetic RF~wave obtained by an oscillating current in a solenoid surrounding the spin system. In this case the Hamiltonian has non-diagonal elements due to the RF~magnetic field $\boldsymbol{\vec{B}_1}=(B_1 cos(\omega t)), B_1 sin(\omega t),0)$ rotating around~$\boldsymbol{\vec{B}_0}$ with an angular velocity~$\omega$:
\begin{align}
\mathscr{H}_{\mathrm{RF}}(t) &= - \frac 12 \gamma \hbar 
\begin{pmatrix}
B_0  &B_1e^{-i\omega t} \\
B_1e^{i\omega t}  & -B_0  \\ 
\end{pmatrix}.
\end{align} 
In this case the time evolution is no longer trivial. We have the following differential equations:
\begin{align}
\begin{split}
\frac{i\hbar \partial\bigl\lvert\psi_{\mathrm{RF}}(t)\bigr\rangle}{\partial t}
&=\mathscr{H}_{\mathrm{RF}}(t)
\begin{pmatrix}
x_2(t) \\
x_1(t) \\ 
\end{pmatrix}\\
&=\tfrac 12\hbar
\begin{pmatrix}
\omega_0  &\omega_1e^{-i\omega t} \\
\omega_1 e^{i\omega t}  & -\omega_0  \\ 
\end{pmatrix}\cdot
\begin{pmatrix}
x_2(t) \\
x_1(t) \\ 
\end{pmatrix},
\end{split}\\
\intertext{where}
\omega_1&=- \gamma B_1.
\end{align}
At resonance, without RF ($B_1=\omega_1=0$) these equations obviously reduce to equation~\eqref{Dif1} and with RF we need to solve the following differential equations:
\begin{align}\label{Dif2}
\begin{pmatrix}
  \dot{x}_2(t)\\
  \dot{x}_1(t)\\
 \end{pmatrix}&=-\frac i2
 \begin{pmatrix}
  \omega_0 x_2(t)+\omega_1e^{-i\omega t} x_1(t)\\
  \omega_1e^{i\omega t} x_2(t)-\omega_0 x_1(t)\\
 \end{pmatrix}.
\end{align}
To solve these two equations we make the following substitutions:
\begin{subequations}
\begin{align}
p(t)&=x_2(t)e^{ i\tfrac{\omega_0}2 t},\\
q(t)&=x_1(t)e^{-i\tfrac{\omega_0}2 t}. 
\end{align}
\end{subequations}
Therefore we have:
\begin{align}
\dot{p}(t)&=\left(-\frac i2\left(\omega_0 x_2(t)+\omega_1e^{-i\omega t} x_1(t)\right)+\frac{i\omega_0}2 x_2(t)\right) e^{i\tfrac{\omega_0}2 t}\nonumber\\
&=-\frac {i\omega_1}2 e^{-i\omega t} x_1(t) e^{i\tfrac{\omega_0}2 t}\\*
&=-\frac {i\omega_1}2 q(t) e^{i(\omega_0-\omega) t} \nonumber\\
\intertext{and}
\dot{q}(t)&=\left(-\frac i2\left(\omega_1e^{i\omega t} x_2(t)-\omega_0 x_1(t)\right)-\frac{i\omega_0}2 x_1(t)\right)e^{-i\tfrac{\omega_0}2 t}\nonumber\\
&=-\frac{i\omega_1}2 e^{i\omega t} x_2(t) e^{-i\tfrac{\omega_0}2 t}\\
&=-\frac{i\omega_1}2 p(t) e^{-i(\omega_0-\omega)t}\nonumber.
\end{align}
When we take the second derivative of $p(t)$ we obtain the following second order differential equation:
\begin{align}
\begin{split}
\label{deriv1}
\ddot{p}(t)&=-\frac{i\omega_1}2\dot{q}(t)e^{i(\omega_0-\omega) t}+i(\omega_0-\omega)\dot{p}(t)\\
&=-\frac{i\omega_1}2 \left(-\frac{i\omega_1}2 p(t) e^{-i\bigl((\omega_0-\omega)t\bigr)}e^{i(\omega_0-\omega) t}\right)+i(\omega_0-\omega)\dot{p}(t)\\
&=i(\omega_0-\omega)\dot{p}(t)-\frac{\omega_1^2}4 p(t).
\end{split}
\end{align}
Let $\lambda_{\pm}$ be the solutions of the equation
\begin{equation}
\label{deriv2}
 \lambda^2-i(\omega_0-\omega)\lambda+\frac{\omega_1^2}4=0.
\end{equation}
We have
\begin{align}
 \lambda_{\pm}=i\frac{\Omega \pm \Delta}2 && \text{with} &&\Omega=\omega_0-\omega && \Delta= \sqrt{\Omega^2+\omega_1^2}
\end{align}
and
\begin{align}
p(t)&=C_1 e^{\lambda_{+}t}+C_2 e^{\lambda_{-} t}, 
\end {align}
therefore:
\begin{subequations}
\begin{align}
\label{x2rf}
\begin{split}
x_2(t)&=p(t)e^{-\tfrac{i\omega_0 t}2}\\
&=\left(C_1e^{\lambda_{+}t}+C_2 e^{\lambda_{-} t}\right)e^{-\tfrac{i\omega_0 t}2}\\
&=e^{-i\tfrac{\omega}2t}\left(C_1 e^{i\tfrac{\Delta}2 t}+
  C_2 e^{-i\tfrac{\Delta}2 t}\right),
\end{split}\\
\intertext{with initial value}
x_2(0)&=r_2 e^{i\phi_2}. 
\end{align}
\end{subequations}
A similar computation for $x_1(t)$ leads to:
\begin{subequations}
\begin{align}
\label{x1rf}
  x_1(t)&=e^{\tfrac{i\omega t}2}\left(C_3 e^{i\tfrac{\Delta}2 t}+
  C_4 e^{-i\tfrac{\Delta}2 t}\right),\\
  x_1(0)&=r_1 e^{ i\phi_1}.
\end{align}
\end{subequations}
As $x_1(t),x_2(t)$ verify the initial value and the differential equations~\eqref{Dif2}
\begin{subequations}
\begin{align}
 C_1+C_2=r_2 e^{i\phi_2},&&
 C_3+C_4=r_1 e^{ i\phi_1},\\*
 (\Delta+\Omega) C_1+\omega_1  C_3=0,&&
 (\Delta-\Omega) C_2-\omega_1  C_4=0, 
\end{align}
\end{subequations}
The computation of the constants~$C_i$ is now straightforward. We have
\begin{subequations}
\label{Ccoef}
\begin{align}
C_1&= \tfrac 1{2}\left((1-\tfrac \Omega{\Delta})r_2 e^{i\phi_2}-\tfrac{\omega_1}{\Delta}r_1 e^{i\phi_1}\right),\\
C_2&= \tfrac 1{2}\left((1+\tfrac \Omega{\Delta})r_2 e^{i\phi_2}+\tfrac{\omega_1}{\Delta}r_1 e^{i\phi_1}\right),\\
C_3&= \tfrac 1{2}\left(-\tfrac{\omega_1}{\Delta}r_2 e^{i\phi_2}+(1+\tfrac \Omega{\Delta})r_1 e^{i\phi_1}\right),\\
C_4&= \tfrac 1{2}\left(\tfrac{\omega_1}{\Delta }r_2 e^{i\phi_2}+(1-\tfrac \Omega{\Delta})r_1 e^{i\phi_1}\right). 
\end{align}
\end{subequations}
We can write the evolution of the spin wave function in matrix notation, therefore equation (\ref{1spinB0}) becomes:
\begin{align}\label {RFmat1}
\bigl\lvert\psi_{\mathrm{RF}}(t)\bigr\rangle&=
\begin{pmatrix}
  x_2(t)\\
  x_1(t)\\
\end{pmatrix}=
E_{RF}(\omega,t) \cdot R_{RF}(\omega,\omega_0,\omega_1,t)\bigl\lvert\psi_{1}(0)\bigr\rangle
\end{align}
The evolution matrix is now dependent on $\omega$ the RF frequency and $\omega_1$ a frequency proportional to the amplitude of RF magnetic field in addition to the time and the magnetic field $B_0$.\\
As shown by equations (\ref{x2rf}),(\ref{x1rf}) and \eqref{Ccoef}, E and R matrices are:
\begin{align}\label {RFmat2}
E_{RF}&=\begin{pmatrix}
e^{-\tfrac{i\omega t}2}&0\\
0&e^{\tfrac{i\omega t}2}\\ 
\end{pmatrix};
&R_{RF}=\begin{pmatrix}
\cos \tfrac{\Delta t}2-\tfrac{i\Omega}{\Delta} \sin \tfrac{\Delta t}2& -i \tfrac{\omega_1 }{\Delta} \sin \tfrac{\Delta t}2\\
 -i \tfrac{\omega_1 }{\Delta} \sin \tfrac{\Delta t}2&\cos \tfrac{\Delta t}2+\tfrac{i\Omega}{\Delta} \sin \tfrac{\Delta t}2\\ 
\end{pmatrix} 
\end{align}
These expressions are modified if the angular velocity~$\omega$ of the RF~magnetic field verifies the resonance condition $\omega=\omega_0$. In that case where $\Omega=0$ and $\Delta=\omega_1$, the evolution matrix becomes $A_{RF}$:
\begin{align}\label{resonance}
A_{RF}(\omega_0,\omega_1,t)&=
\begin{pmatrix}
e^{-\tfrac{i\omega_0 t}2}&0\\
0&e^{\tfrac{i\omega_0 t}2} 
\end{pmatrix}\cdot
\begin{pmatrix}
\cos\tfrac{\omega_1 t}2&-i \sin\tfrac{\omega_1 t}2\\[0.5em]
-i \sin\tfrac{\omega_1 t}2&\cos\tfrac{\omega_1 t}2 
\end{pmatrix}.
\end{align}
Therefore the frequency reference is the same for $\omega_0$  and $\omega$ which gives an experimental way to define a precise time scale with its origin given by the electrical current used in the RF coil.\\
In magnetic resonance experiments the effect of a RF~pulse is usually described~\cite{coh} as a rotation with an angle of $\theta_1=\omega_1 t$ in the spin space around the vector $\boldsymbol{\vec{u}}=\left(u_x,u_y,u_z \right)$ as:
\begin{align}
\boldsymbol{R_{u,\theta_1}^{\tfrac 12}}&=
\begin{pmatrix}
\cos\tfrac{\theta_1}2-iu_z \sin\tfrac{\theta_1}2 & \left( -iu_x-u_y\right) \sin\tfrac{\theta_1}2 \\[0.5em]
\left( -iu_x+u_y\right)\sin\tfrac{\theta_1}2 & \cos\tfrac{\theta_1}2+iu_z \sin\tfrac{\theta_1}2 
\end{pmatrix}.
\end{align}
In NMR $\boldsymbol{\vec{u}}$ lies in the transverse plane: $\boldsymbol{\vec{u}}=\left( 1, 0, 0 \right)$ where the $x$-direction is given by the phase of the RF pulse (in other words by the beginning of the RF pulse) and therefore rotations induced in NMR are described by:
\begin{align}
\label{eq5}
\boldsymbol{R_{(1,0,0),\theta_1}^{\tfrac 12}}&=
\begin{pmatrix}
\cos\tfrac{\theta_1}2 & -i\sin\tfrac{\theta_1}2 \\[0.5em]
-i\sin\tfrac{\theta_1}2 & \cos\tfrac{\theta_1}2 
\end{pmatrix}.
\end{align}
We notice that this is exactly the transformation matrix that we have computed at resonance in equation  \eqref{resonance}, except that the time evolution $E(\omega_0,t)$ is missing in this equation. Therefore in the general case, equation \eqref{RFmat1} gives the evolution of the wave function and equation \eqref{RFmat2} shows that the evolution of the spin system is a time evolution given by matrix $E(\omega,t)$ which depends on the RF frequency followed by a rotation of angle $\Delta t= \sqrt{(\omega_0-\omega)^2+\omega_1^2}\ t$ around the vector:
\begin{equation}
\boldsymbol{\vec{u}_\omega}=\left( u_x=\dfrac{\omega_1} {\sqrt{(\omega_0-\omega)^2+\omega_1^2}},\ u_y=0,\ u_z =\dfrac{\omega_0-\omega} {\sqrt{(\omega_0-\omega)^2+\omega_1^2}}\right). 
\end{equation}
It is important to note that matrix product is non-commutative. In other words, the spin evolution in presence of a RF field is, according to equations \eqref{RFmat1} and \eqref{RFmat2}, a matrix product of a time evolution operator and a rotation operator. The wave function after the pulse is given by:             
\begin{align}
\label{1spinRF}
\bigl\lvert\psi_{\mathrm{RF}}(t)\bigr\rangle&=A_{RF}\cdot\bigl\lvert\psi_{1}(0)\bigr\rangle
=E_{RF}\cdot R \cdot  \bigl\lvert\psi_{1}(0)\bigr\rangle
\end{align}
which is the same as the equation derived in Abragam textbook \cite{abragam} but here we have calculated explicitly the $A_{RF}$ matrix. Therefore at resonance the wave function at the end of the RF pulse is equal to the wave function multiplied by the rotation matrix of angle  $\theta_1= \omega_1 t$ which is then multiplied by the evolution matrix but there is no time sequence between these two operations. This elementary description of the spin dynamics retrieve the main characteristics of NMR experiments, but it is not complete.
\subsection{The other part of the Hamiltonian.}
Usually the Hamiltonian is developed only for the part of the system under investigation. For example the hydrogen atom in a magnetic field is studied for the energy levels of the electron without any reference to the proton or the quarks. The effect of the magnetic field on the proton is usually seen as a second order perturbation of the electron spin system. In H nuclear magnetic resonance, the studied system is the proton but the magnetic field has an effect on the electron and on the proton. For several hydrogen atoms the kinetic part of each proton is dependent on the structure of the spin system and the kinetic energy of the proton ($p^2/2m$) has to be considered. More precisely the kinetic energy is $\frac {-\hbar^2}{2m} \nabla^2(F(\mathbf {r},t)) $ where F is a function of space and time dependent on the spin system. For our purpose we consider the energy of the hydrogen atom due to the proton kinetic energy and to the nuclear forces to be very great compared to the magnetic-spin interaction. This energy will be called $K'$. The time fluctuations of $K'$ are very small, it is why the fluctuations are usually neglected in the evolution of the Hamiltonian $\mathscr{H}_{T}(t)$ but it can not be canceled because it has important effects as it  will be seen through this work. Hence, for a spin system that experiences an external  $\boldsymbol{\vec{B}_0}$~ the complete Hamiltonian for one spin$\tfrac 1 2$  is defined by:
\begin{align}
\mathscr{H}_{T}(t) & =K' -\vec{\mu} \cdot \vec{B}_0\\ \nonumber
&=\tfrac 12  \hbar
\begin{pmatrix}
\omega_0 +\frac{2K'}{ \hbar} &\frac{2K'}{ \hbar} \\
\frac{2K'}{\hbar}  & -\omega_0+\frac{2K'}{\hbar}  \\ 
\end{pmatrix}
\end{align}
and  the Schr\"{o}dinger equation becomes:
\begin{align}\label{Dif2T}
\begin{pmatrix}
  \dot{x}_{2T}(t)\\
  \dot{x}_{1T}(t)\\
 \end{pmatrix}&=-\frac i2
 \begin{pmatrix}
 ( \omega_0+K) x_{2T}(t)+K x_{1T}(t)\\
K x_{2T}(t)-(\omega_0-K)  x_{1T}(t)\\
 \end{pmatrix}
\end{align}
with $K=2K'/ \hbar$. We call the $K$ parameter measured in hertz the rest constant.\\
Here to solve these equations we make the following substitutions:
\begin{align}
\label{kinetic1}
p(t)&=x_2(t)e^{i \tfrac{\omega_0 +K}{2}t},
&& q(t)=x_1(t)e^{-i\tfrac{\omega_0 -K}{2}t}. 
\end{align}
which gives if $\omega_0$ and $K$ are time independent:
\begin{align}
\label{kinetic2}
\ddot{p}(t)-i\omega_0\dot{p}(t)+\frac{K^2}4 p(t)=0
&&\ddot{q}(t)+i\omega_0\dot{q}(t)+\frac{K^2}4q(t)=0
\end{align}
which are the same equation as that obtained in equation \eqref{deriv1} with $ \Omega=\omega_0$ and $\omega_1=K$. The solution of these equations is:
\begin{align}
\begin{pmatrix}
C_1 e^{i\tfrac{\Delta -K}{2}t}+  C_2 e^{-i\tfrac{\Delta+K}{2}t}\\[0.8em]
C_3 e^{i\tfrac{\Delta -K}{2}t}+  C_4 e^{-i\tfrac{\Delta +K}{2}t}
\end{pmatrix}
&=\begin{pmatrix}
e^{-i \tfrac{K}{2}t}&0\\
0&e^{-i \tfrac{K}{2}t}\\
\end{pmatrix}\cdot
\begin{pmatrix}
C_1 e^{i\tfrac{\Delta}{2}t}+C_2 e^{-i\tfrac{\Delta}{2}t}\\[0.8em]
C_3 e^{i\tfrac{\Delta }{2}t}+C_4 e^{-i\tfrac{\Delta }{2}t} 
\end{pmatrix}
\end{align}
where 
$C_1, C_2,C_3,C_4$ are identical to the coefficients obtained in the previous case given by equations~\eqref{Ccoef} where $\Delta$ is $\sqrt{\omega_0^2+K^2}$ and exchanging $\Omega$ with $ \omega_0$ and $\omega_1$ with $K$.
This solution is again the product of an evolution matrix with a rotation matrix:
\begin{align}\label{wfkinetic}
\bigl\lvert\psi_{\mathrm{T}}(t)\bigr\rangle&=E_T \cdot  R_T \bigl\lvert\psi_{1}(0)\bigr\rangle
\end{align}
with
\begin{align}\nonumber
E_{T}&=\begin{pmatrix}
e^{-\tfrac{i K t}2}&0\\
0&e^{-\tfrac{i K t}2}\\ 
\end{pmatrix};
& R_{T}=\begin{pmatrix}
\cos \tfrac{\Delta t}2-\tfrac{i\omega_0}{\Delta} \sin \tfrac{\Delta t}2& -i \tfrac{K }{\Delta} \sin \tfrac{\Delta t}2\\
 -i \tfrac{K }{\Delta} \sin \tfrac{\Delta t}2&\cos \tfrac{\Delta t}2+\tfrac{i\omega_0}{\Delta} \sin \tfrac{\Delta t}2\\ 
\end{pmatrix} 
\end{align}
Therefore the evolution of the wave function of one spin is given by the product of the spin evolution operator and a spin rotation of angle $\theta =\sqrt{\omega_0^2+K^2} \: t$ around $\vec{u}(K/\Delta,0, \omega_0/\Delta)$. The evolution operator induces no oscillation between the two states therefore it seems that no magnetic resonance signal can be detected but we will see that if the magnetic field is not homogeneous the spin motion introduces oscillations between the states even without RF magnetic fields.\\
It is important to notice that the rest constant $K$ of the Hamiltonian, has a tremendous effect on the spin motion in a homogeneous magnetic field because there is no more $\omega_0$ in the $E_{T}$ evolution part of the wave function given by equation \eqref{wfkinetic}.
\subsection{General description of one spin$\tfrac 12$}
In the general case the Hamiltonian of a spin$\tfrac 12$ is dependent on the time evolution of the magnetic field $B(t)$, therefore it can be written as:
\begin{align}\label{gen1}
\mathscr{H}_1 & =-\tfrac 12 \hbar
\begin{pmatrix}
\gamma B_Z -K &\gamma B_X -i \gamma B_Y -K \\
\gamma B_X + i \gamma B_Y -K & -\gamma B_Z-K  
\end{pmatrix}
\end{align}
and the Schr\"{o}dinger equation becomes:
\begin{align}\label{Dif2Tb}
\begin{pmatrix}
  \dot{x}_{2}\\
  \dot{x}_{1}\\
 \end{pmatrix}&=-\frac i2
 \begin{pmatrix}
 ( \omega_Z+K) x_{2}+(\omega_X-i\omega_Y+K) x_{1}\\
(\omega_X+i\omega_Y+K) x_{2}-(\omega_Z-K)  x_{1}
 \end{pmatrix}
\end{align}
with $K=2 K'/ \hbar$,  $ \omega_X=-\gamma B_X$,  $\omega_Y=-\gamma B_Y$ ,  $\omega_Z=-\gamma B_Z $ . This equation gives the spin motion whatever the magnetic fields not only with an important magnetic field in the Z direction direction as stated in conventional NMR description.\\ 
We look for a constant solution by an adequate exponential substitution $p(t)=x_2(t)e^{A(t) }$ and  $q(t)=x_1(t)e^{B(t)}$ which their derivatives give:
\begin{subequations}
\label{general1}
\begin{align}
\dot{p}&= p \left[ \dot{A} - \tfrac i2(\omega_Z+K) \right] - \tfrac i2 ( \omega_X -i\omega_Y+K) q e^{(A-B)}\\
\dot{q}&= q \left[ \dot{B} + \tfrac i2(\omega_Z-K) \right] - \tfrac i2 ( \omega_X +i\omega_Y+K) p e^{(B-A)}
\end{align}
\end{subequations}
 If we take the parameters at one time $A=\tfrac i2(\omega_Z+K)t$, $B=-\tfrac i2(\omega_Z- K)t$,  one obtains:
\begin{subequations}
\label{general3}
\begin{align}
\ddot{p}&-i \omega_Z \dot{p}+\tfrac 14(\omega_X^2+\omega_Y^2+K^2+2K\omega_X)p=0  \\ 
\ddot{q}&+i \omega_Z \dot{q}+\tfrac 14(\omega_X^2+\omega_Y^2+K^2+2K\omega_X)q=0 
\end{align}
\end{subequations}
In case of constant magnetic field $B(0,0,B_0)$ we retrieve equations \eqref{kinetic2} but the comparison with these equations show that the general derivation introduces a coupling of the  rest constant and the X component of the magnetic field in case of non constant magnetic fields. The solutions of equations \eqref{general3} is, as usually, obtained by the constant solution then varying the constants.\\
The solution of these equations, equivalent to equations (24-25), is:
\begin{subequations}
\label{general4}
\begin{align}
x_2(t)&=p(t)e^{-A(t)} =e^{-\tfrac i2(\omega_Z+K)t} \left(C_1 e^{\lambda_{+}t}+C_2 e^{\lambda_{-} t}\right)\\
x_1(t)&=q(t)e^{-B(t)} =e^{\tfrac i2(\omega_Z-K)t} \left(C_3 e^{-\lambda_{-}t}+C_4 e^{-\lambda_{+} t}\right)
\end{align}
\end{subequations}
where
\begin{align}
\label{general5}
 \lambda_{\pm}=\frac{i\left(\omega_Z\pm \Delta\right)}2,
  && \Delta= \sqrt{\omega_Z^2+\omega_1^2} ,
  && \omega_1^2= \omega_X^2+\omega_Y^2+K^2+2K\omega_X\\
    \omega_Z=\omega_0 -K
\end{align}
The $C_i$ coefficients are given by equations \eqref{Ccoef}  where $\Omega$ is substituted by $\omega_Z$ ($\Omega =\omega_0-\omega=\omega_Z$).\\
This substitution is not only a mathematical tool but it has the physical significance that the off resonance of the RF is equivalent to a magnetic field oscillation in the magnetic field direction which is the Z direction. \\
The general evolution of the spin wave function is:
\begin{align}
\label{generalw}
\bigl\lvert\psi(t)\bigr\rangle&=\begin{pmatrix}
  x_2(t)\\
  x_1(t)\\
\end{pmatrix}=
E(\omega_0,K,t) \cdot R(\omega,\omega_0,\omega_1,t) \cdot \bigl\lvert\psi_{1}(0)\bigr\rangle,
\end{align}
The solution with the parameters used for the differential equation resolution is:
\begin{align}\label{general6}
\bigl\lvert\psi(t)\bigr\rangle&=
\begin{pmatrix}
x_2(t)\\
x_1(t)
\end{pmatrix}=
\begin{pmatrix}
e^{- \frac{i}{2} (\Omega +K)t}&0\\
0&e^{ \frac{i}{2} (\Omega -K)t}
\end{pmatrix}
\cdot
\begin{pmatrix}
a  & b\\
b & \bar{a}  
\end{pmatrix}
\cdot
\begin{pmatrix}
r_2 e^{i\phi_2}\\
r_1 e^{i\phi_1}
\end{pmatrix}
\end{align}
with
\begin{align} \nonumber
&a=\cos \tfrac{\Delta t}2-\tfrac{i\Omega}{\Delta} \sin \tfrac{\Delta t}2
&&b= -i \tfrac{\omega_1 }{\Delta} \sin \tfrac{\Delta t}2
\end{align}
This equation shows that evolution of the spin states between the two states exists if it exists a magnetic field of intensity about K ($\Omega=\omega_Z$ in the general case). If $\Omega=0$ the two states have the same time evolution $e^{- \frac{i}{2} K t}$. We will see that the presence of another spin with its magnetic moment $\boldsymbol{\vec{\mu}}=\frac 12 \gamma \hbar\boldsymbol{\vec{\sigma}}$ will always introduce a local magnetic field.
\subsection{Two spins $\tfrac 12$}
For two spins we consider a four dimensional Hilbert space which is a tensor product of the two dimensional Hilbert spaces of the one spin system. We will distinguish two cases: one for an homogeneous magnetic field~$\boldsymbol{\vec{B}_0}$ and one where the two spins are in different magnetic fields.
\subsubsection{Two spins in an homogeneous magnetic field}
In the case of an homogeneous magnetic field, each spin experiences the same magnetic field and the Hamiltonian of the system transforms a 2-dimensional Hilbert space in 4-dimensional Hilbert space. In order to preserve the scalar product, the Kronecker sum is used. Kronecker sum is defined by:
\begin{align} 
A \oplus_K B = A\otimes I_n + I_m \otimes B,
\end{align} 
with $\otimes$ the tensor product with identity matrices $I$ of dimensions $n$ the first dimension of matrix $B$ and $m$ the second dimension of matrix $A$, (the important property of Kronecker sum is that $e^{A \oplus_K B}= e^A \otimes e^B$).\\
For matrices 2x2 one obtains:
\begin{align}
A=
\begin{pmatrix}
            a  &  b \\
	         c  & d  \\
           \end{pmatrix},&&
B=
\begin{pmatrix}
             \alpha  & \beta \\
	     \gamma & \delta  \\
           \end{pmatrix},
\end{align}

\begin{align} 
A \oplus_K B =
\begin{pmatrix}
            a+ \alpha  &b  & \beta & 0 \\
            c  & d+ \alpha  & 0 & \beta \\
            \gamma  & 0  & a+\delta & b \\	     
            0 & \gamma & c  & d+\delta \\	     
           \end{pmatrix}
\end{align}
The Hamiltonian of two spin$\tfrac12$ in an homogeneous magnetic field is therefore:
\begin{align} \label{homo2}
\mathscr{H}_{2,\mathrm{ Hom}}=-\tfrac12  \gamma\hbar
\bigl[(\sigma_1^x\oplus_K\sigma_2^x\bigr)B_X+\bigl(\sigma_1^y\oplus_K\sigma_2^y\bigr)B_Y+\bigl(\sigma_1^z\oplus_K\sigma_2^z\bigr)B_Z\bigr],
\end{align}
with
\begin{align}\label{Kpauli}
\sigma_p^x=\begin{pmatrix}
            K_p  &  K_p+1 \\
	     K_p+1  &  K_p  \\
           \end{pmatrix},&&
\sigma_p^y=\begin{pmatrix}
             K_p  &  K_p -i \\
	     K_p +i &  K_p  \\
           \end{pmatrix},
\end{align} 
and
\begin{align}          
\sigma_p^z=\begin{pmatrix}
            K_p +1 &  K_p \\
	   K_p & K_p-1  \\
           \end{pmatrix}.\nonumber
\end{align}
\begin{align}\label{Kp}
 K_p=-2K'_p/ \hbar
\end{align}
 $K_p=-K/ \omega_p$ is dimensionless as shown in $\mathscr{H}_{2,\mathrm{ Hom}}$ through the matrices $(\sigma_p^x, \sigma_p^y, \sigma_p^z)$, the minus sign indicates that energy reference is given by the spin system,  $\omega_p$ is the Larmor frequency for each $p$ nucleus measured in the range of tens to hundreds of megahertz in frequency space~\cite{rya} for high resolution NMR. $K'_p$ has the same dimension as $\hbar$. Note that the matrices are no more Pauli matrices but are still the components of a vector $\boldsymbol{\vec{\sigma}_p}$.\\
In very homogeneous magnetic field where $\boldsymbol{\vec{B}}=(0,0,B)$  and  $B_X=B_Y=0$ the Hamiltonian \eqref{homo2} can be written:
\begin{align}\label{homoB}
&\mathscr{H}_{2,\mathrm{B}} =  - \frac 12 \gamma \hbar \\ \nonumber
&\begin{pmatrix}
(K_1+K_2+2)B & K_1 B & K_2 B  & 0 \\
K_1 B & (K_1+K_2)B & 0  & K_2 B \\
K_2B  & 0 & (K_1+K_2)B &K_1 B \\
0  & K_2 B  & K_1 B & (K_1+K_2-2)B 
\end{pmatrix}
\end{align}
This equation shows that for these two spins the rest constant of the  Hamiltonian becomes important by introducing a coupling between the spins even without any component in the orthogonal plane to the magnetic field direction. The wave function can be obtained by the eigenvalues of the Hamiltonian which is no more a tensor product of the individual wave functions because the kinetic part is dependent on the position of the spin.
 The important thing is that the two spins are no more a linear combination of the two spin states but a four spin state which can be considered as entangled spin states.
\subsubsection{Two spins in different magnetic fields}
If the magnetic fields experienced by spins~$A$ and~$B$ are respectively~$\boldsymbol{\vec{B}_A}$ and~$\boldsymbol{\vec{B}_B}$, then the Hamiltonian of the system is:
\begin{align}\label{difB1}
\mathscr{H}_2&=-\left((\boldsymbol{\vec{\mu}_A} \cdot \boldsymbol{\vec{B}_A})\oplus_K(\boldsymbol{\vec{\mu}_B} \cdot \boldsymbol{\vec{B}_B})\right).
\end{align}
The two spins are now distinguishable by their magnetic environment which is always greater than the magnetic field heterogeneity and $K_A=K_B=0$ because the kinetic part can be included in the local  magnetic field. If we write this Hamiltonian in matrix form, we obtain the following:
\begin{align}
\mathscr{H}_2&= \!
-\frac {\gamma \hbar}2\!  
\begin{pmatrix}
B_{AZ}+B_{BZ}  & B_{BX}-iB_{BY} & B_{AX}-iB_{AY}  & 0 \\
B_{BX}+iB_{BY}  & B_{AZ}-B_{BZ} & 0  & B_{AX}-iB_{AY}\\
B_{AX}+iB_{AY}  & 0 & B_{BZ}-B_{AZ}  & B_{BX}-iB_{BY} \\
0  & B_{AX}+iB_{AY} & B_{BX}+iB_{BY}  & -B_{AZ}-B_{BZ}  
\end{pmatrix}\!\!,
\end{align} 
If we suppose that both magnetic fields are in the $Oz$-direction, this matrix reduces to:
\begin{align}\label{Ham2Oz}
\mathscr{H}_2&= 
- \frac {\gamma \hbar}2  
\begin{pmatrix}
B_A+B_B  & 0 & 0  & 0 \\
0  & B_A-B_B & 0  & 0\\
0  & 0 & B_B-B_A  & 0 \\
0  & 0 & 0  & -(B_A+B_B)  
\end{pmatrix}.
\end{align}
We directly obtain the eigenvalues of the Hamiltonian~\eqref{Ham2Oz}:
\begin{subequations}
\begin{align}
E_{++} &= \frac12\gamma \hbar (B_A+B_B),\\
E_{+-} &= \frac12\gamma \hbar (B_A-B_B),\\
E_{-+} &= -\phantom{-}\frac12\gamma \hbar (B_A-B_B),\\
E_{--}&=- \phantom{-}\frac12\gamma \hbar (B_A+B_B). 
\end{align}
\end{subequations}
This leads to the following time-dependent wave function:
\begin{align}\label{4wavefun}
\bigl\lvert\psi_2(t)\bigr\rangle &= 
\begin{pmatrix}
r_4e^{-\tfrac i2\bigl((\omega_A+\omega_B)t-\phi_{4}\bigr)} \\
r_3e^{-\tfrac i2\bigl((\omega_A-\omega_B)t-\phi_{3}\bigr)} \\
r_2e^{ \tfrac i2\bigl((\omega_A-\omega_B)t-\phi_{2}\bigr)} \\
r_1e^{ \tfrac i2\bigl((\omega_A+\omega_B)t-\phi_{1}\bigr)}  
\end{pmatrix},
\end{align}
where:
\begin{subequations}
\begin{align}
\omega_A&=-\gamma B_A,\\
\omega_B&=-\gamma B_B.
\end{align}
\end{subequations}
The wave function for two independent spins can be written as a tensor product of the wave function of each spin:
\begin{equation}\label{tensor2spin}
\bigl\lvert\psi_2(t)\bigr\rangle=\bigl\lvert\psi_1(t)\bigr\rangle_{\omega_A}\!\!\otimes\bigl\lvert\psi_1(t)\bigr\rangle_{\omega_B},
\end{equation}
This is simply done by taking equation~\eqref{tensor2spin} as the proper definition for the case of two spins~$\tfrac12$ and using equation~\eqref{1spinB0} for each single wave function to obtain the following wave function for two spins:
\begin{align}
\begin{split}
\bigl\lvert\psi_2(t)\bigr\rangle&=\bigl\lvert\psi_1(t)\bigr\rangle_{\omega_A}\!\!\otimes\bigl\lvert\psi_1(t)\bigr\rangle_{\omega_B}\\
&=
\begin{pmatrix}
r_{A2}e^{-i\bigl(\tfrac{\omega_At}{2} +\phi_{A1}\bigr)}\\
r_{A1}e^{i\bigl(\tfrac{\omega_At}{2} +\phi_{A1}\bigr)} 
\end{pmatrix}\otimes
\begin{pmatrix}
r_{B2}e^{-i\bigl(\tfrac{\omega_Bt}{2} +\phi_{B2}\bigr)}\\
r_{B1}e^{i\bigl(\tfrac{\omega_Bt}{2} +\phi_{B1}\bigr)}
\end{pmatrix}\\
&=
\begin{pmatrix}
r_4e^{-i\bigl(\tfrac{(\omega_A+\omega_B)t}{2} +\phi_{4}\bigr)}\\
r_3e^{-i\bigl(\tfrac{(\omega_A-\omega_B)t}{2} +\phi_{3}\bigr)}\\
r_2e^{i\bigl(\tfrac{(\omega_A-\omega_B)t}{2} +\phi_{2}\bigr)}\\
r_1e^{i\bigl(\tfrac{(\omega_A+\omega_B)t}{2} +\phi_{1}\bigr)}
\end{pmatrix}.
\end{split}
\end{align}
with
\begin{align}
r_1&=r_{A1}r_{B1}	&\phi_{1}=\phi_{A1}+\phi_{B1} \\ \nonumber
r_2&=r_{A2}r_{B1}	&\phi_{2}=\phi_{A2}+\phi_{B1} \\ \nonumber
r_3&=r_{A1}r_{B2}	&\phi_{3}=\phi_{A1}+\phi_{B2} \\ \nonumber
r_4&=r_{A2}r_{B2}	&\phi_{4}=\phi_{A2}+\phi_{B2}  \nonumber
\end{align}
We see that we retrieve the fundamental rule:
\begin{align}
\sum r_i^2=1
\end{align}
only if the phase of each spin is distinguishable $\phi_A$ or $\phi_B$. In this case the two spins are not in an entangled state.
\subsubsection{Two spins in different magnetic fields with a RF field}
Using the tensor product notation we can directly compute the wave function for two spins in an inhomogeneous magnetic field with an RF~magnetic field~$\boldsymbol{\vec{B}_1}$ rotating around~$\boldsymbol{\vec{B}_0}$ with an angular velocity~$\omega$:
\begin{align} \label{RFgen}
\bigl\lvert\psi_2(t)\bigr\rangle&=\bigl\lvert\psi_1(t)\bigr\rangle_{\omega_A}\!\!\otimes\bigl\lvert\psi_1(t)\bigr\rangle_{\omega_B}\nonumber\\
&=\Bigl(E(\omega)\cdot R(\omega_A)\cdot\bigl\lvert\psi_1(0)\bigr\rangle_{\omega_A}\Bigr)\otimes\Bigl(E(\omega)\cdot R(\omega_B)\cdot\bigl\lvert\psi_1(0)\bigr\rangle_{\omega_B}\Bigr)\\*
&=\bigl(E(\omega)\otimes E(\omega)\bigr)\cdot\bigl(R(\omega_A)\otimes R(\omega_B)\bigr)\cdot\Bigl(\bigl\lvert\psi_1(0)\bigr\rangle_{\omega_A}\!\!\otimes\bigl\lvert\psi_1(0)\bigr\rangle_{\omega_B}\Bigr).\nonumber
\end{align}
\subsubsection{Two spins$\tfrac 12$ in the general case}
If the rest constant of the Hamiltonian is not neglected, the eigenvalues of the Hamiltonian are no more reduced to no off diagonal components but are given by the Kronecker sum of the Hamiltonian \eqref{gen1} and the wave function is given by solving an extension to matrices (4X4) of the differential equations \eqref{general6}. 
\subsection{N spins$\tfrac 12$}
The description we have obtained for the Hamiltonian of two spins can be generalized for a $N$~spin system by decomposing it in q parts of homogeneous magnetic field $\boldsymbol{\vec{B_q}}(B_{qX},B_{qY},B_{qZ})$ with $n_q$  spins in each homogeneous part. The Hamiltonian of the spins in each spatial domain q is given by the Kronecker sum of the  $\sigma$ three components:
\begin{align} \label{homoNq}
\mathscr{H}_{n_q,\mathrm{ Hom}}=-\tfrac12\gamma\hbar
\left( ( \bigoplus_{p=1}^{n_q}\sigma_p^x\bigr)B_{qX}+(\bigoplus_{p=1}^{n_q}\sigma_p^y\bigr)B_{qY}+(\bigoplus_{p=1}^{n_q}\sigma_p^z\bigr)B_{qZ}\right),
\end{align}
The $\sigma$ components are still defined by the \eqref{Kpauli} equations. Therefore this Hamiltonian can be understood as it has the dot product of a new magnetic moment $\boldsymbol{\vec{\mu}_q}$ and the local magnetic field $\boldsymbol{\vec{B}_q}$ with  $n_q$ spins in each domain.\\
The Hamiltonian of the N=$\sum_1^q n_q$ spins  is given by the generalization of equation \eqref{difB1}:
\begin{align}\label{difN}
\mathscr{H}_N =-\left(\bigoplus_{q=1}^{q}(\boldsymbol{\vec{\mu}_q} \cdot \boldsymbol{\vec{B}_q})\right).
\end{align}\\
In each homogeneous magnetic field $B_q$ where there is only one wave function corresponding to $n_q$ entangled spins obey the Fermi-Dirac statistics for spin$\tfrac 1 2$. The q domains which correspond to $\boldsymbol{\vec{\mu}_q}$ distinguishable magnetic moments, obey the Boltzmann statistics currently used for NMR experiments explanations. 
\section{NMR experiments}
A way to test this theoretical description of  spin evolution is to look at proton NMR which is a very good example for studying isolated spins. 
\subsection{The nuclear spin Hamiltonian}
 If we consider one Hydrogen nucleus its total energy is $m^{'} c^2$ where  $m^{'}$ is the mass of the moving nucleus and its magnetic energy $-\mu B$ is tremendously low compared to the total energy. In an homogeneous magnetic field the Hamiltonian is given by Equations \eqref{homo2} for two spins and equation \eqref{homoNq} for N spins. \\
In NMR the applied magnetic field is far greater than local magnetic fluctuations arising from electron spin and space becomes oriented by an homogeneous magnetic field which by convention is the z-axis. There is another   more implicit convention which is the change of the reference frame which introduces the gyromagnetic ratio and a change of the origin of the energy scale. The origin of the energy scale becomes by the Larmor frequency of the spin under investigation no more an independent physical constant. $K_p$ defined by equation \eqref{Kp} and  used in equations \eqref{homoB} and \eqref{homoNq} can be understood as the ratio that contributes in the part of magnetic energy from the total energy. The Hamiltonian for $N$ nuclear spins is given by an extended matrix based on the equation \eqref{homoB} where for $K_p=0$ only the energy of the first and the last of the $2^N$ states are non zero because the intermediary states cancel two by two. Therefore only these  two states have a time evolution: $r_1 e^{i(\omega_0 t+ \phi_1)}$ and $r_{2^N} e^{-i(\omega_0 t- \phi_{2^N})}$ but there is no transition between these two states because only $r_1^2+r_2^2=1$ and $r_{2^N}^2+r_{2^N-1}^2=1$ are states of the same spin. Therefore there is no evolution of the spin states and we retrieve the important experimental fact that there is no NMR signal at equilibrium.
\subsection{Complex susceptibility of one spin$\tfrac 12$}
 In NMR, the measurement of the spin system is obtained by the current induced in a tuned $LCR$ circuit. The influence of the sample is determined from its complex susceptibility $\chi = \chi'(\omega)+i\chi"(\omega)$ which expresses the linear relationship between magnetization and magnetic field \cite{slich2}. It is well known that after an adequate $RF$ pulse an electrical current is induced in the $LCR$ circuit.\\
For one spin we propose to define the complex susceptibility by:
\begin{align}\label{measure}
\chi=\gamma \hbar  \left( \bar{x}_1 (t) x_2(t) \right),
\end{align}
where $\bar{x}_1 (t)$ is the complex conjugate of the amplitude of state $\left\lvert-\right\rangle$ and $x_2(t)$ the amplitude of state $\left\lvert+\right\rangle$ of the wave function of the spin. As the amplitude of the states are related by $\lvert x_1(t)\rvert^2+\lvert x_2(t)\rvert^2=1$ the relative phase information is given by the complex conjugate of one of the two amplitudes. In a static magnetic field $B_0$ very homogeneous and without the rest constant $K$, the wave function is given by the conventional equation \eqref{1spinB0} where $x_1(t)=r_1 e^{i(\tfrac {\omega_0 t}{2}+\phi_1)}$, $x_2(t)=r_2 e^{-i(\tfrac {\omega_0 t}{2}-\phi_2)}$and the definition \eqref{measure}, equal to
\begin{align} \label{M10}
\chi_0(t)=\gamma\hbar r_1 r_2 e^{-i(\omega_0t +\phi_1-\phi_2)}
\end{align}
In the general case we have to use matrix \eqref{general6} describing the spin wave evolution and the complex susceptibility becomes:
\begin{align} \label{M1t}
\chi(t)=\gamma\hbar e^{-i\Omega t}
 \left\lbrace 
 (a\: \bar{b}\: r_2^2+a\:b\:r_1^2)
 +r_1 r_2(b\:\bar{b}\:e^{i(\phi_1-\phi_2)}+a^2e^{i(\phi_2-\phi_1)})
\right\rbrace 
\end{align}
with
\begin{align}\label{first term}
(a\: \bar{b}\: r_2^2+a\:b\:r_1^2)=
 \tfrac{\omega_1}{\Delta} (\tfrac{\Omega}{\Delta} sin^2 \tfrac{\Delta t}{2} +i \cos \tfrac{\Delta t}{2} \sin \tfrac{\Delta t}{2}) (r_2^2-r_1^2)= D_1 (r_2^2-r_1^2)
\end{align}
and 
\begin{align}\label{second term}
r_1 r_2\left(  \tfrac{\omega_1^2}{\Delta^2} sin^2\tfrac{\Delta t}{2} e^{i(\phi_1-\phi_2)}+( cos^2\tfrac{\Delta t}{2}-\tfrac{\Omega^2 }{\Delta^2} sin^2\tfrac{\Delta t}{2}-2i\tfrac{\Omega}{\Delta}  \cos \tfrac{\Delta t}{2} \sin \tfrac{\Delta t}{2} )e^{i(\phi_2-\phi_1}\right)  
\end{align}
and
\begin{align}
 \Omega=\omega_Z ;
  && \omega_1^2= \omega_X^2+\omega_Y^2+K^2+2K\omega_X ;
  && \Delta= \sqrt{\omega_Z^2+\omega_1^2} 
\end{align}
These equations show that  K of the Hamiltonian is included in the $\omega_1$ parameter. The susceptibility of one spin can be written as:
\begin{align}\label{chione}
\chi(t)=\gamma\hbar\: e^{-i \: \Omega \: t}
 \left\lbrace 
 D_1 (r_2^2-r_1^2) +r_1 r_2 ( D_2\: e^{i(\phi_1-\phi_2)} + D_3 \: e^{i(\phi_2-\phi_1)}
\right\rbrace
\end{align}
For N independent spins $\tfrac 12$, we can consider two cases the first one where the rest constant K is not taken into account (K=0) and the second case where K is present in the Hamiltonian of the spin set.
\subsection{Low field NMR experiments}
Magnetic resonance experiments was, at the beginning of the NMR story, performed with homogeneous magnetic fields of about 0.1T  with an homogeneity better than $\Delta B_0 /B_0 = 10^{-5}$. By convention the Z direction of the reference frame is given by the direction of the magnetic field. As the gyro-magnetic ratio of the proton is $\gamma= 2.7\times 10^8\: rad\: s^{-1}\: T^{-1}$. According to  $\omega_Z=- \gamma B$, for  a magnetic field of 1T and an homogeneity of $10^{-5}$, the angular velocities are typically $\omega _Z=- 2.7\times 10^8 rad\: s^{-1}$ and  $2.7\times 10^3\: rad\: s^{-1}$ for $\omega _X$ and $\omega _Y$. For a 3 $cm^3$ water sample there is 1/3 mole of water which corresponds to $N=2 \times 10^{24}$ protons.\\
As there is no NMR signal at equilibrium, the way to induce it is to use a RF magnetic field. In this case the complex susceptibility of N spins is the sum of the susceptibility of one spin given by equation \eqref{chione} :
\begin{align}\label{sucepdis}
\chi_{N}(t)&=\sum _{q=1}^{q=N}\int_{-\pi}^{\pi}  \chi_q(t) d\phi_q
\end{align}
The difference $\phi_q=\phi_{2q}-\phi_{1q}$ is random but is defined for each spin therefore the integral over $\phi_q$ is null and the total susceptibility of N independent spins is:
\begin{align}\label{suscepG}
\chi_{N}(t)&=\sum _{q=1}^{q=N} \int_{-\pi}^{\pi} \chi_q(t) d\phi_q=\gamma\hbar N  e^{-i\:\Omega \:t} D_1 \sum _{q=1}^{q=N}(r_{2q}^2-r_{1q}^2)
\end{align}
with
\begin{align} \nonumber
D_1= \tfrac{\omega_1}{\Delta} (\tfrac{\Omega}{\Delta} \sin^2 { \tfrac{\Delta t}{2}} +i \cos { \tfrac{\Delta t}{2}}\sin { \tfrac{\Delta t}{2}}) 
\end{align}
The parameters are:
\begin{align}
\label{par RF}
 \Omega=\omega_Z
  && \Delta= \sqrt{\omega_Z^2+\omega_1^2} ,
  && \omega_1^2= \omega_X^2+\omega_Y^2+K^2+2K\omega_X
\end{align}
In NMR experiments, the RF magnetic field  $\omega_Y=\omega_X$ used corresponds to a 1 mT oscillating magnetic field with an amplitude much more than the proton K fluctuations and about 1000 times less than the main magnetic field. The $\omega$ RF frequency used is precisely at the resonance frequency: $\omega=\omega_0=\omega_Z$ and with K equal to 0 because all the magnetic fields present are far greater than the K parameter fluctuations, therefore $(\omega_1=\sqrt{2}\omega_X$ and $\Delta=\omega_0$). With these parameters the complex susceptibility of the N spins becomes:
\begin{align}\label{NMR RF}
\chi_{N}(t)&=\gamma\hbar \: e^{-i\omega_0 t}\tfrac{\sqrt{2}\omega_X}{\omega_0} \left[ \sin^2 { \tfrac{\omega_0 t}{2}}+i \tfrac {1}{2} \sin \omega_0 t \right]  N \sum _{q=1}^{q=N}(r_{2q}^2-r_{1q}^2)\\ \nonumber
&=\gamma\hbar \:\tfrac{\sqrt{2} \omega_X}{2\omega_0}\left[e^{-i\omega_0 t} - \:e^{-i \:2\omega_0  t}\right]   N \sum _{q=1}^{q=N}(r_{2q}^2-r_{1q}^2)
\end{align}
where $N \sum (r_{2q}^2-r_{1q}^2)$ corresponds to the Boltzmann distribution of the N spins in the N first $2^N$ spin states. This equation retrieves the less known experimental fact that there are two resonances which can be detected one at $\omega_0$ and another at $2\omega_0$ \cite{abragam}. The NMR signal is maximum when the pulse has a duration $t_p$ such as $\omega_X t_p=\pi /2$ corresponding to the so called $\pi/2$ pulse.
\subsection{High resolution NMR}
High resolution NMR is characterized by very homogeneous high magnetic fields typically $\Delta B_0 / B_0=  10^{-8}$. In this case the magnetic field experienced by one spin becomes dependent on the orientation of neighborhood spins and the Hamiltonian is given by equation \eqref{difB1}. If all the magnetic fields are in the same direction the Hamiltonian is given by equation \eqref{Ham2Oz}. For example in  ethanol ($CH_3-CH_2-OH$) solution  there are three groups of equivalent protons: the 3 protons of the $CH_3$ methyl group, the two protons of the $CH_2$ group and the proton of the OH group. Schematically, the protons of the methyl group experience a magnetic local field $\omega_{Zl}$ dependent on the orientation of the neighborhood two spins. The local magnetic fields are $\omega_{Zl}=\omega_Z+\epsilon$ if the two spins are in (Z) direction, $\omega_{Zl}=\omega_Z$ if one spin is in (Z) direction and the other in (-Z) direction this combination is twice more probable than the preceding one and $\omega_{Zl}=\omega_Z-\epsilon$ if both spins are in (-Z) direction. Therefore the NMR signal characterized by the amplitude of each $\omega_{Zl}$ frequencies is composed of 2 neighborhood protons +1. Therefore for the protons of the methyl group of ethanol molecules  there are 3 frequencies with relative amplitude (1,2,1). The wave function with a RF field is according to equation \eqref{RFgen} the tensor product of the evolution matrix of each spin and the tensor product of the rotation matrices. Therefore the NMR spectrum of the methyl group is still three resonances with the same relative amplitudes (1,2,1). Again the amplitude of the NMR signal is maximum after a $\pi/2$ pulse.
\subsection{Nuclear spin noise experiments}

Felix Bloch  \cite{Bloch1946} has predicted the theoretical possibility of an emission from nuclear spins without any magnetic field. The first observation of random noise emission from nuclear spins was performed by Sleator, Hahn, Hilbert and Clarke \cite{Hahn85, Hahn87} using a superconducting SQUID. At very low temperature  without applied magnetic field the other part of the Hamiltonian becomes important and can explain an emission from nuclear spins via the spin motion in the magnetic fields generated by the neighborhood spins and can be measured.\\ 
In very homogeneous magnetic fields where $\Delta B_Y /B_0=10^{-8}$ the rest constant of the Hamiltonian has also to be taken into account and can explain nuclear spin emission without any RF magnetic field even at room temperature. MCCoy and Ernst \cite{Ernst1989} have observed nuclear spin noise emission at room temperature using a conventional (300 MHz) NMR spectrometer. They compared the proton resonance of the methyl group in ethanol without any RF pulse just by co-adding 8000 measurements with the proton resonance obtained with a conventional one $\pi/2$ RF pulse. The most  surprising feature of the spectrum without any RF is the negative deviation compared to the spectrum obtained by a $\pi/2$ excitation pulse and secondly an amplitude of the signal which is about $10^8$ time smaller. Our derivation retrieves these main characteristics.\\
The wave function without RF is given by an extension to N spins of equation \eqref{general6} where the spin system for the proton of the methyl protons is composed of three local magnetic fields. The parameters of equation \eqref{suscepG} which gives the complex susceptibility of N spin is $\Omega=\omega_0$, $\omega_1= K$ and $\Delta=\omega_0$ therefore the complex susceptibility is:
\begin{align}\label{NMR no RF}
\chi_{noise}(t)=\gamma\hbar \:\tfrac{K}{2\omega_0}\left[e^{-i\omega_0 t} - \:e^{-i \:2\omega_0  t}\right]   N \sum _{q=1}^{q=N}(r_{2q}^2-r_{1q}^2)
\end{align}
which is the same as equation \eqref{NMR RF} for experiment with RF except where K is replaced by $\sqrt{2} \omega_X$. The rest constant can be expressed in frequency unit according to the substitution given by equation \eqref{kinetic1}.\\
In NMR, $\omega_X$ is always positive because all the frequency measurements are locked to the RF signal by the use of lock-in amplifiers, without such a locking tool the phase of the electrical current becomes random and no accumulation can be performed. As the kinetic part of the rest constant is proportional to $-\hbar/m_p$ the complex susceptibility with no RF is always negative. The amplitude of the complex susceptibility for a defined number of spin is given by $ \sqrt 2 \omega_X/ \omega_0$ with RF and $ K/ \omega_0$ without RF. As K is about $6\times 10^{-8} rad/s$ for one proton it seems not measurable but K is a scalar and in homogeneous magnetic field of $\Delta B/B= 10^{-8}$ a part of the N spins have the same rest constant and the complex susceptibility becomes measurable. The possibility to measure a NMR signal is always dependent on the homogeneity of the magnetic field even in low magnetic field  NMR experiments.
\subsection{NMR imaging without RF experiments}
\subsubsection{External lock experiments}
In an imaging experiment, where magnetic gradient fields are applied in all space directions M\"{u}ller and Jerschow \cite{ Muller2006} have obtained an image without RF field, showing that entangled spins can be modified by magnetic field gradients. They used a 500 MHz chemical spectrometer with magnetic field gradients of 6 mT/m strength oriented in direction orthogonal to $B_0$ for $X,Y=0^o$ to $X,Y=180^o$ by step of $6^o$. Each step was measured 512 times. To prevent any leakage or accidentally delivery RF power to the probe, the RF power transmission cable was disconnected. The obtained image is a way to visualize the spatial distribution of the rest constant. Again the synchronization of NMR signals was obtained by the external lock of high resolution NMR spectrometers which gives the time origin and $\omega_X$.
\subsubsection{ Experiments with no external lock}
 Usually NMR imaging is obtained by a first gradient applied in one direction with a RF pulse in order to select a slice characterized by its position on this first direction and its thickness given by resonance condition. After that a gradient is applied in an orthogonal direction in order to give a phase coherence and thereafter a gradient is applied in the third orthogonal direction during the signal acquisition. In a previous work \cite{Kees 2014}, NMR images were obtained using such method with a very small RF power. In fact a very small RF power is automatically applied in the used spectrometer with no external lock. In the experiment described by \cite{Kees 2014} the RF power was 1 $\mu$W  which for a 10 cm diameter coil, is less than the electrical noise level in such a coil. Therefore image in Figure 2B of \cite{Kees 2014} can be considered as an image obtained without RF showing that conventional NMR imaging sequence can be used for spin noise imaging if the signal acquisition is properly synchronized. Another experimental fact showing the importance of synchronization is that the MCCoy and Ernst experiment \cite{Ernst1989} is not reproducible on a Bruker 7T BS70/30 system without any external lock for synchronization. 
\subsection{Summary}
 In summary we analyzed different experimental effects that show the following  physical implications of the solutions of Schr\"{o}dinger's equation with the complete Hamiltonian:
\begin{itemize}
\item We retrieve for the low field NMR experiments the less known experimental fact that there are two detectable resonances one at $\omega_0$ and  another at $2 \omega_0$. Furthermore, the NMR signal is maximum when the pulse has a duration corresponding to the so called $\pi/2$ pulse which shows that time unit is dependent on the studied NMR frequency.
\item In the case of high resolution NMR for the protons of the methyl group of ethanol molecules there are 3 frequencies with the relative amplitudes (1,2,1), and the maximum amplitude of the NMR signal is still obtained after a $\pi/2$ pulse.
\item Our derivation retrieves the main characteristics of the spectrum without any RF, i.e. the nuclear spin noise experiments for the phase change and the NMR signal amplitudes.
\item NMR imaging without RF experiments has been discussed in both cases with and without external locks showing the experimental importance to define a zero time of acquired signals in order to sum them adequately.
\end{itemize}
\section{ Conclusion}
This work has shown that an elementary approach of the quantum description of an ensemble of independent spin $\tfrac12$ can retrieve the magnetic resonance basic experimental facts by an adequate definition of the complex susceptibility. This derivation completes the density matrix formalism which is well adapted for describing NMR experiments where spin coupling by magnetic fields induced by neighborhood spins have measurable effects. Experiments using nuclear spin noise are difficult to be explained by the matrix density formalism without introducing an extra probability factor and a stochastic operator as in the paper of Field and Bain \cite{field}.\\
The present work has proposed that in very homogeneous magnetic field the non external magnetic part of the Hamiltonian becomes important by introducing a coupling between the spins even without any magnetic component in the plane orthogonal to the magnetic field direction. In this case the spins$\tfrac12$ into the homogeneous magnetic field are in an entangled state  which can be used for quantum computing. NMR is still a good example of an experimental quantum computer and the use of magnetic gradients can be a way of increasing the number of qubits. The main result of our approach is to predict that the quantum interference effects are best detected in a very homogeneous magnetic field where many spins are in entangled spin states. Magnetic field gradients can be used to initiate the system and the RF pulses can be used for creating logical gates and for measuring the result of quantum entanglements computation \cite{Qcomp}. NMR images without RF is the proof that encoding gradients can select a voxel where the magnetic field is sufficiently homogeneous that it can be considered as an entangled NMR quantum bit. Previous NMR experiments on quantum computing  have shown that NMR qubits can be manipulated in order to create quantum gates \cite{QNMR 2000,Qcomp nature 2001}. Using magnetic field gradients can tremendously increase the number of qubits compared to molecule based qubits and can give a new way to do quantum computing by NMR especially at low temperature where the loss of coherence effects are reduced.

\end{document}